\documentclass[journal,10pt]{IEEEtran}
\usepackage[T1]{fontenc}
\usepackage{cite}
\usepackage{flushend}
\usepackage{amsmath,graphicx}
\usepackage{amssymb}
\usepackage{algpseudocode}
\usepackage{amsfonts}
\usepackage{graphicx,subfigure}
\usepackage{fancyhdr}  %
\usepackage{cases}
\usepackage{extarrows}
\usepackage{algorithm}
\usepackage{multirow,tabularx}
\usepackage{mathtools}
\usepackage{xcolor}
\usepackage[english]{babel}
\usepackage{caption}
\usepackage{autobreak}
\usepackage{bm}

\makeatletter
\renewcommand{\maketag@@@}[1]{\hbox{\m@th\normalsize\normalfont#1}}%
\makeatother
\ifCLASSINFOpdf
\else
\fi

\begin{document}
\title{Low-Complexity Beam Training for Multi-RIS-Assisted Multi-User Communications\vspace{-1mm}}
\author{Yuan Xu, Chongwen Huang,~\IEEEmembership{Member,~IEEE}, Li Wei, Zhaohui Yang, Xiaoming Chen,~\IEEEmembership{Senior Member,~IEEE}, \\ 
Zhaoyang Zhang,~\IEEEmembership{Senior Member,~IEEE}, Chau Yuen,~\IEEEmembership{Fellow,~IEEE}, and \\
Mérouane Debbah,~\IEEEmembership{Fellow,~IEEE}\vspace{-7mm}
\thanks{Y. Xu, C. Huang are with College of Information Science and Electronic Engineering, Zhejiang University, Hangzhou 310027, China, with the State Key Laboratory of Integrated Service Networks, Xidian University, Xi’an 710071, China, and Zhejiang Provincial Key Laboratory of Info. Proc., Commun. \& Netw. (IPCAN), Hangzhou 310027, China (E-mails: yuan\_xu@zju.edu.cn, chongwenhuang@zju.edu.cn). \\
\indent X. Chen, Z. Yang and Z. Zhang are with College of Information Science and Electronic Engineering, Zhejiang University, Hangzhou 310027, China, and also with Zhejiang Provincial Key Laboratory of Info. Proc., Commun. \& Netw. (IPCAN), Hangzhou 310027, China (E-mails: \{chen\_xiaoming, zhaohui\_yang, ning\_ming\}@zju.edu.cn).\\
\indent L. Wei and C. Yuen are with the School of Electrical and Electronics Engineering, Nanyang Technological University, Singapore 639798 (E-mails: l\_wei@ntu.edu.sg, chau.yuen@ntu.edu.sg).\\
\indent M. Debbah is with the Department of Electrical Engineering and Computer Science and the KU 6G Center, Khalifa University, Abu Dhabi 127788, UAE, and also
with CentraleSupelec, University Paris-Saclay, 91192 Gif-sur-Yvette, France (E-mail: Merouane.Debbah@ku.ac.ae).\\
\indent The work was supported by the China National Key R\&D Program under Grant 2021YFA1000500 and 2023YFB2904800, National Natural Science Foundation of China under Grant 62331023, 62101492, 62394292, 62231009 and U20A20158, Zhejiang Provincial Natural Science Foundation of China under Grant LR22F010002, Zhejiang Provincial Science and Technology Plan Project under Grant 2024C01033, Zhejiang University Global Partnership Fund, and Singapore Ministry of Education (MOE) Academic Research Fund Tier 2 MOE-T2EP50220-0019.  
}}
\maketitle
\thispagestyle{empty}
\pagestyle{empty}

\begin{abstract}
In this paper, we investigate the beam training problem in the multi-user millimeter wave (mmWave) communication system, where multiple reconfigurable intelligent surfaces (RISs) are deployed to improve the coverage and the achievable rate. However, existing beam training techniques in mmWave systems suffer from high complexity (i.e., exponential order) and low identification accuracy. To address these problems, we propose a novel hashing multi-arm beam (HMB) training scheme that reduces the training complexity to the logarithmic order with the high accuracy. Specifically, we first design a generation mechanism for HMB codebooks. Then, we propose a demultiplexing algorithm based on the soft decision to distinguish signals from different RIS reflective links. Finally, we utilize a multi-round voting mechanism to align the beams. Simulation results show that the proposed HMB training scheme enables simultaneous training for multiple RISs and multiple users, and reduces the beam training overhead to the logarithmic level. Moreover, it also shows that our proposed scheme can significantly improve the identification accuracy by at least 20\% compared to the existing beam training techniques.
\end{abstract}
	


\vspace{-3.2mm}
\section{Introduction}\label{sec:intro}
\vspace{1.3mm}
Reconfigurable intelligent surfaces (RISs) have attracted rapidly growing interest\cite{Di9140329}, mainly due to their potential to enhance spectrum/energy efficiency, mitigate interference, and enable manual customization of wireless communication environments. Specifically, RISs are two-dimensional metamaterial antenna arrays composed of a large number of inexpensive elements. These elements can be dynamically controlled to manipulate the radio propagation environment, facilitating the establishment of virtual line-of-sight (LoS) links between base stations (BSs), RISs, and users\cite{Huang8741198,10077561}.
\par
In millimeter wave (mmWave) communications, the signals suffer from significant path loss. Large array antenna and narrow beams with RISs techniques are commonly used to compensate for the signal degradation. Before data transmission, the beam training process plays a crucial role in aligning the transmitter and receiver beams to establish reliable links and maximize the received power\cite{You9129778}. However, the accuracy of beam alignment for narrow beams in millimeter waves is demanding. Furthermore, the deployment of multiple RISs significantly increases the multiplex gain that also needs to be aligned.
\par
\textcolor{black}{In recent years, various beam training techniques and algorithms have been proposed, including the exhaustive beam training\cite{Junyi5262295}, hierarchical beam training\cite{Li8751142,Oh8451870}, equal interval multi-arm beam (EIMB) training method\cite{You9129778}, and two-timescale-based beam training\cite{10106469}.} The exhaustive beam training method searches all possible beam directions at both the transmitter and receiver\cite{Junyi5262295}, resulting in significant delays and exponential complexity\cite{Wang9771330}. The hierarchical beam training method employs a multi-stage approach, dividing the beam space into two halves at each stage until the desired resolution is achieved, which offers a lower complexity but suffers from inherent drawbacks. That is, using wide beams in early layers reduces the beamforming gain, leading to identification errors. Moreover, these errors accumulate at subsequent subdivided beam layers\cite{Hur6600706}. The EIMB training method employs a predetermined codebook and gradually narrows down the search space through multiple rounds of training until it finds the aligned direction\cite{You9129778}. However, it depends on the results of the first round and the fixed beam composition method introduces leakage interference that is difficult to eliminate, which may limit the accuracy of beam identification in complex situations to some extent. Further, existing methods cannot support simultaneous training of multiple transmitters or receivers and instead need to take turns in a sequential manner, resulting in sub-optimal performance.
\par
To address these problems, in this paper, we propose a novel beam training method that exhibits low complexity as well as high accuracy. Specifically, we consider the uplink multi-RIS-assisted multi-user mmWave communication system. \textcolor{black}{Beam training on the BS side is not mentioned due to space limitations, but hashing beam training is applicable to arbitrary multi-antenna arrays. Taking the RIS-user link as an example, our method works as follows: In each time slot, each user transmits a pilot signal, and we construct the receive hashing multi-arm beams at the RISs. The multiple RISs then reflect the signals to the BS. Assuming that the BS can distinguish the signals of different users from the received superimposed signal power, we design a demultiplexing algorithm based on the soft decision and a multi-round voting mechanism, to determine the aligned beams of different RISs to users. It's worth noting that we choose independent hash functions for each RIS to generate multi-arm beams, which ensures the minimal correlation between different RIS reflection links. }
\par
Furthermore, the randomness of hash functions, the soft decision, and the multi-round voting design can improve the identification accuracy. Then, the proposed method significantly reduces the training complexity, since it allows for simultaneous training of multiple RISs and uses multi-arm beams as well. Simulation results show the outstanding performance of our proposed beam training method in terms of both high identification accuracy and low training overhead compared to existing methods.

\vspace{-2.5mm}
\section{System Model}\label{sec:format}
\vspace{0.3mm}
\captionsetup[figure]{name={Fig.}}
\begin{figure}[t]
	\begin{center}
		\centerline{\includegraphics[width=0.35\textwidth]{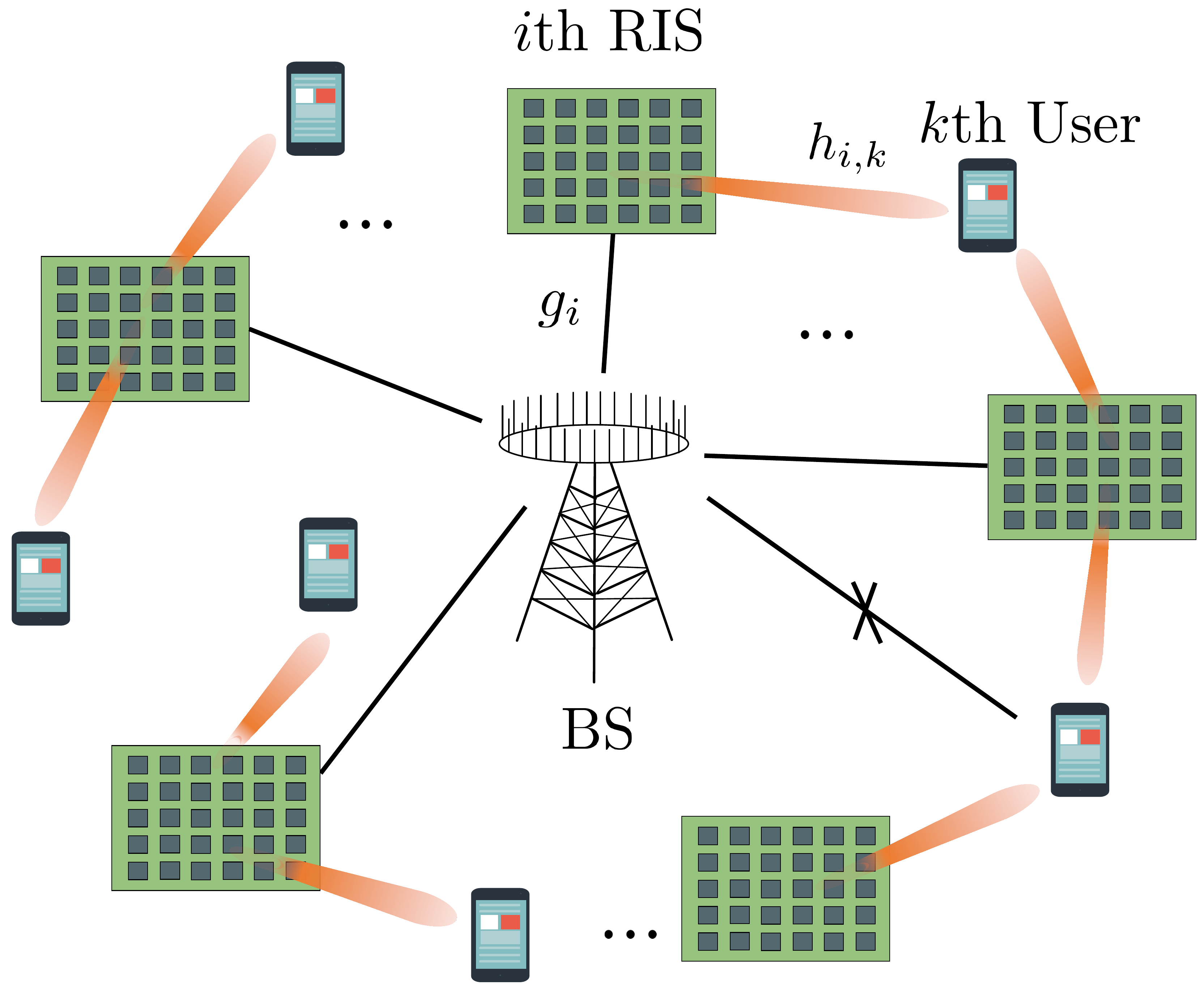}}  \vspace{-1mm}
		\caption{The considered multi-RIS-assisted multi-user mmWave communication system, consisting of a $N_A$-antenna BS, $K$ single-antenna mobile users, and $I$ distributed RISs.  }
		\label{fig:System_Scheme} 
	\end{center}\vspace{-8mm}
\end{figure} 
In this section, we introduce the system model for the considered uplink multi-RIS-assisted multi-user mmWave communication. As illustrated in Fig.~\ref{fig:System_Scheme}, the direct links between the $N_A$-antenna BS and the cluster of $K$ single-antenna users are blocked, and $I$ distributed RISs are deployed to assist the transmission. Each RIS, denoted by $i \in \mathcal{I}=\left\{1,2,\ldots,I\right\}$, consists of $N_i$ reflective elements and maintains LoS connections with both the BS and each user. 
\par
\textcolor{black}{The signal transmitted by the user $k$ and reaching the BS by the reflection of RISs in $\mathcal{I}$, can be expressed as
\begin{equation}
	\begin{split}\label{con:yk}
		y_{k}=\sum\limits_{i\in\mathcal{I}}\sqrt{P}\mathbf{v}_k\left(\mathbf{g}_{i}^H diag(\bm{\theta}_i) \mathbf{h}_{i,k}\right)x_k+n_{i,k},
	\end{split}\vspace{-1mm}
\end{equation}
where $x_k\in\mathbb{C}$ denotes the signal transmitted by the $k$-th user with transmission power $P$ and $n_{i,k}\sim\mathcal{CN}(0,\sigma_n^2)$ denotes the additive white Gaussian noise. The effective channel from the $i$-th RIS to the BS and from the $k$-th user to the $i$-th RIS are denoted as $\mathbf{g}_{i}$ and $\mathbf{h}_{i,k}$ respectively, and $\mathbf{v}_k$ is the BS beamforming vector for the $k$-th user. The reflection vector of the $i$-th RIS is defined as $\bm{\theta}_i\triangleq [e^{j\psi_{1}},e^{j\psi_{2}},\ldots,e^{j\psi_{N_i}}]\in\mathbb{C}^{1\times N_i}$, where $\psi_{n_i}$ denotes the phase shift at the $n$-th element.}
\par
In particular, mmWave communication channels typically follow the geometric model\cite{Ayach6717211}. Therefore, the channels $\mathbf{g}_{i}$ and $\mathbf{h}_{i,k}$ can be expressed as
\vspace{-1mm}
\begin{equation} \mathbf{g}_{i}=g_{i}\mathbf{a}(\phi_{i}^t,\theta_{i}^t)\mathbf{b}(\phi_{i}^r,\theta_{i}^r),
\end{equation}
\begin{equation}
	\mathbf{h}_{i,k} =h_{i,k}\mathbf{a}\left(\phi_{i,k}^r,\theta_{i,k}^r\right),
\end{equation}
where $g_{i}$ denotes the complex gain of the RIS $i$-BS channel, and $h_{i,k}$ denotes the complex gain of the user $k$-RIS $i$ channel. $\phi_{i}^t$ and $\theta_{i}^t$ ($\phi_{i}^r$ and $\theta_{i}^r$) represent the azimuth angle and the elevation angle at the RIS (BS) respectively for the RIS $i$-BS link; $\phi_{i,k}^r$ and $\theta_{i,k}^r$ represent the azimuth angle and the elevation angle at the RIS for the user $k$-RIS $i$ link. The array steering vectors associated with the RISs and the BS are represented by $\mathbf{a}$ and $\mathbf{b}$, respectively. 
\par
For a uniform planar array (UPA) with $N_i=N_h\times N_v$ antennas, where $N_h$ and $N_v$ represent the number of horizontal and vertical array elements, the steering vector can be expressed as\vspace{-1mm}
\begin{equation}\label{equ:YXZ}
	\begin{split} \mathbf{a}&(\phi,\theta)=\left[1,e^{j\frac{2\pi}{\lambda}d_v\sin\theta},\ldots,e^{j\frac{2\pi}{\lambda}d_v(N_v-1)\sin\theta}\right] \\
&\otimes\left[1,e^{j\frac{2\pi}{\lambda}d_h\sin\phi \cos\theta},\ldots,e^{j\frac{2\pi}{\lambda}d_h(N_h-1)\sin\phi \cos\theta}\right],
	\end{split}
\end{equation}
where $\mathbf{a}\in\mathbb{C}^{1\times N_i}$, $d_h$ and $d_v$ denote the horizontal array element spacing and the vertical array element spacing, respectively. $\lambda$, $\phi$, and $\theta$ denote the wavelength of the signal, the azimuth angle, and the elevation angle respectively, and $j=\sqrt{-1}$ is the imaginary unit.
\vspace{-2mm}
\section{Hashing Beam Training}\label{sec:beam_alignment}
\vspace{1.8mm}
\subsection{Generation Mechanism of Hashing Multi-Arm Beams}
\vspace{1.4mm}
Firstly, we introduce the single-beam generation process. Generally, by discretizing the azimuth angle $\phi\in[0,\pi]$ and the elevation angle $\theta\in [0,\pi]$ in the three-dimensional (3D) space into $N_1$ and $N_2$ directions respectively, and substituting into the UPA steering vector defined by Eq.~(\ref{equ:YXZ}), a single-beam codebook $\mathbf{C}\in\mathbb{C}^{N\times N_i}$ can be designed as Eq.~(\ref{eq:C})\cite{You9129778}. The $n_1N_2+n_2+1$-th row of the codebook $\mathbf{C}$ represents the codeword that corresponds to a beam covering the spatial region defined by $\sin\phi \cos\theta\in[\frac{2n_1}{N_1}-1,\frac{2n_1+2}{N_1}-1]$, $\sin\theta\in[\frac{2n_2}{N_2}-1,\frac{2n_2+2}{N_2}-1]$, where $n_1\in [0,...,N_1-1]$, and $n_2\in [0,...,N_2-1]$.
\begin{figure*}[t]
\centering\vspace{-2mm}
    \begin{equation}\label{eq:C}
	\begin{split}
		\mathbf{C}=&[\mathbf{a}(\phi_1,\theta_1);...;\mathbf{a}(\phi_1,\theta_{N_2});...\mathbf{a}(\phi_{N_1},\theta_1);...;\mathbf{a}(\phi_{N_1},\theta_{N_2})]\vspace{-2mm}
	\end{split}
    \end{equation}
    \begin{equation}\label{eq:F}
	\bm{\mathcal{F}}=
	\setlength{\arraycolsep}{1pt}
	\begin{bmatrix}
		1&1&...& 1\\
		1&\xi_h^{f_h}\xi_v^{f_v}&...&\xi_h^{f_h(N_h-1)}\xi_v^{f_v(N_v-1)}\\
		\vdots&\vdots&\ddots&\vdots\\
		1&\xi_h^{f_h(N_1-1)}\xi_v^{f_v(N_1-1)}&...&\xi_h^{f_h(N_1-1)(N_h-1)}\xi_v^{f_v(N_2-1)(N_v-1)}
	\end{bmatrix}
    \end{equation}
    \begin{subequations}\label{eq:s}
        \begin{equation}
            \tilde{\mathbf{s}}=[\mathbf{s}_1;\ldots;\mathbf{s}_B]^T
        \end{equation}
    \begin{equation}\vspace{-7mm}
	\begin{split}
	\mathbf{s}_b=[&\mathbf{C}(d_b(1),1\!:\!M),\mathbf{C}(d_b(2),1\!+\! M\! :\!2M),...,\mathbf{C}(d_b(R),(R\!-\!1)M\!+\!1\! :\!N_i)]
	\end{split}
    \end{equation}
\end{subequations}
{\noindent} \rule[-10pt]{17.5cm}{0.05em}\vspace{-3mm}\\
\end{figure*}
\begin{figure}[t]\vspace{.5mm}
	\begin{center}
		\centerline{\includegraphics[width=0.5\textwidth]{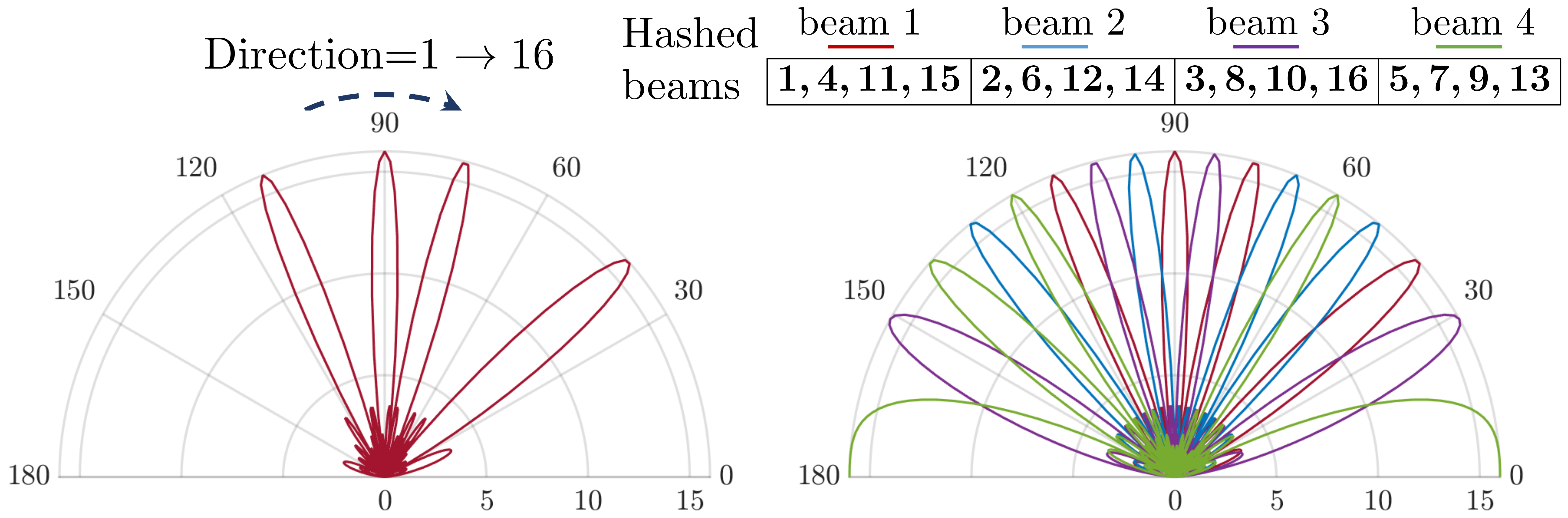}}  \vspace{-1mm}
		\caption{Illustration of the proposed multi-arm beams generating method when $N_1=16$, $N_2=1$, $B=4$.  }
		\label{fig:2D} 
	\end{center}\vspace{-8mm}
\end{figure} 
\par 
By substituting $\sin\phi \cos\theta=f_h$ and $\sin\theta=f_v$, we obtain a standard two-dimensional (2D) discrete Fourier transform (DFT) matrix of size $N_1N_2\times N_hN_v$ as Eq.~(\ref{eq:F}), where $\xi_h=e^{\frac{-j2\pi d_h}{\lambda}}$, and $\xi_v=e^{\frac{-j2\pi d_v}{\lambda}}$. That is, the antenna units resemble the time sample, and the signal directions resemble the frequency.
\par
\textcolor{black}{Compared to the single-beam training method, our training method utilizes multi-arm beams, significantly reducing the number of training beams in the entire search space. Specifically, for an arbitrarily chosen hash function $f:\mathcal{U}\to\mathcal{T}$, we map $N=N_1N_2$ directions into $B$ multi-arm beams, where $\mathcal{U}=\{0,1,...,N-1\}$ and $\mathcal{T}=\{0,1,...,B-1\}$. The direction collection of all multi-arm beams can be represented as $\tilde{\mathbf{d}}=[\mathbf{d}_1;...;\mathbf{d}_B]^T\in\mathbb{C}^{B\times R}$. The $b$-th multi-arm beam contains $R = N/B$ sub-directions, denoted as $\mathbf{d}_b=[d_b(1),...,d_b(R)]\in\mathbb{C}^{1\times R}$, where $d_b(r)\in\{1,...,N\}$ denotes the $r$-th sub-direction. The hash function generation method is presented in the supplementary material due to space constraints. For the sake of understanding, we plot the hashing multi-arm beams (HMB) in Fig.~\ref{fig:2D}, where $16$ directions are hashed by a hash function into $4$ multi-arm beams. The left plot is the first multi-arm beam, which contains $4$ sub-directions, and it can be seen that they are well dispersed.}

\par
To construct the corresponding codeword $\mathbf{s}_b$ of the multi-arm beam $\mathbf{d}_b$, we segment the antenna array and use each partition to generate a sub-beam in the multi-arm beam. Specifically, we truncate $M=N_i/R$ adjacent elements from the $d_b(r)$-th row of the single-beam codebook $\mathbf{C}\in\mathbb{C}^{N\times N_i}$. Then we splice them up as in Eq.~(\ref{eq:s}), where $\mathbf{C}(d_b(r),m_1\! :\!m_2)$ denotes the elements in the $d_b(r)$-th row, columns $m_1$ to $m_2$ of the codebook $\mathbf{C}$. As a result, the $r$-th partition of $M$ adjacent elements corresponds to a beam pointing in the $d_b(r)$-th single-beam direction, and the beam width is increased by a factor of $R$ compared to the single beams. 
\par
Based on the generation mechanism of the HMB and the corresponding codebook, beam training can be performed on the RIS side. The beam training process consists of two major phases, the scanning phase and the beam identification phase which are elaborated as follows.
\vspace{-4.3mm}
\subsection{Scanning Phase}
\vspace{0.3mm}
To find the direction $(\phi^*,\theta^*)$ of users relative to the RISs, our beam training consists of $L$ rounds of scanning. In each round of scanning, we randomly select a hash function for each RIS from an independent hash function family. Specifically, for the $i$-th RIS, we randomly choose a hash function to generate the multi-round hashing multi-arm beams $\mathbf{D}^i=[\tilde{\mathbf{d}}^i_1;...;\tilde{\mathbf{d}}^i_L]$ and the corresponding codebook $\mathbf{S}^i=[\tilde{\mathbf{s}}^i_1;...;\tilde{\mathbf{s}}^i_L]$, where $\tilde{\mathbf{d}}^i_l$ and $\tilde{\mathbf{s}}^i_l$ denote the multi-arm beam and the codeword in the $l$-th round of scanning. This ensures that each round of scanning covers the entire beam space, providing a comprehensive exploration of potentially aligned beam directions. 
\par
In each time slot of the uplink, all users transmit training symbols once and the receive vector $\bm{\theta}_i$ of the $i$-th RIS is selected from the predefined codebook $\mathbf{S}^i$ until all the codewords have been traversed. This means that multiple RISs can generate different multi-arm beams to reflect the signals from users simultaneously. Note that the selection order of codewords is unimportant. For example, in the $q=(l-1)B+b$-th time slot, the $i$-th RIS can use the codeword $\mathbf{s}^{i,l}_b$ as the receive steering vector. Thus, user signals are reflected by multiple RISs, and then superimposed and recorded at the BS. 
\vspace{-2.7mm}
\subsection{Beam Identification Phase}
\vspace{1.mm}
We assume that the BS is capable of distinguishing signals from different users using existing techniques\cite{Hara642841,Armstrong4785281,Yin4534773}. For the sake of illustration, we will focus on a typical user, since the beam identification phase operates the same for each user. Suppose that the direction of the typical user with respect to RIS $i$ is denoted as $\gamma_i\in[1,...,N]$. We can represent the received signal power vector over $Q=LB$ time slots as $\mathbf{P}^r=[P^r(1),...,P^r(Q)] \in\mathbb{C}^{Q\times 1}$, where $P^r(q)$ denotes the superimposed signal power received by the BS at the $q\in\mathbf{q}=\{1,...,Q\}$-th time slot from the typical user. In the $q$-th time slot, the RIS may or may not maximally align its beam with the user, which we call seeing the user. We denote the situation where the $i$-th RIS can align its beam with the user at time slot $q$, i.e., $\gamma_i\in \mathbf{D}^i(q,:)$, as $e^i_q=1$. On the other hand, we denote $\gamma_i\notin \mathbf{D}^i(q,:)$ as $e^i_q=0$. Obviously, the number of times when $e^i_q=1$ is equal to the number of scan rounds $L$, since each direction is assigned to a multi-arm beam in each round.
\par
The designed beam identification phase incorporates a demultiplexing algorithm based on the soft decision and a multi-round voting mechanism. The demultiplexing algorithm extracts valuable information from the power of the superimposed signals, which is dependent on two important facts. Firstly, the hash functions used to generate the hashing multi-arm beams for different RISs are independent of each other. This ensures that the correlation between different RIS reflection links is minimized. Consequently, the probability of two arbitrary RISs aligning their beams with the user in the same time slot simultaneously can be expressed as
\vspace{-1mm}
\begin{equation}
	 Pr(e^i_q\underset{(i\neq j)}{=} e^j_q=1) = 1/B^2.\vspace{-1.7mm}
\end{equation}
\par
\textcolor{black}{For example, when $B=4$, we have the probability $Pr(e^i_q\underset{(i\neq j)}{=} e^j_q=1)= 1/16$, which is sufficiently small. Thus, it can be guaranteed that the useful component of the received signal is reflected by at most one RIS as long as the number of multi-arm beams $B>I$. It is worth noting that when $I$ is relatively large, we can ensure that $B>I$ always holds by designing a larger $B$, or by grouping the RIS and then training each group in turn.}
\par
Secondly, due to the varying attenuation of RIS reflective links, there are different reflected channel gains and signal strengths. This allows us to rank the reflected signal strengths as $P_{m_1}>P_{m_2}>...>P_{m_I}$, where $m_i$ corresponds to the RIS reflective link with the $i$-th strongest reflected channel gain, and $P_{m_i}$ represents the corresponding signal power. Accordingly, we can employ the soft decision to identify the useful signal powers reflected by different RISs. Specifically, we assign the time slots with the $(i-1)L\!+\!1$-th to $iL$-th strongest received signal power in vector $\mathbf{P}^r$ to RIS $m_i$, which means\vspace{-2mm}
\begin{subequations}
\setlength\abovedisplayskip{-1pt}
\setlength\belowdisplayskip{-1pt}
    \begin{equation}
	\begin{split}
		 \mathbf{q}_{m_i}=\arg\max\limits_{(i-1)L+1:iL} descend(\mathbf{P}^r),
	\end{split}
\end{equation}
\begin{equation}\vspace{-2mm}
	\begin{split}
		e^i_{q^*}=1,\ q^*\in\mathbf{q}_{m_i},
	\end{split}
\end{equation}
\end{subequations}
where $descend(\cdot)$ represents sorting the vector in descending order.
\par
Now that we have identified the $L$ time slots for each RIS that contain user signals, we can proceed with the multi-round voting mechanism to find the direction $\gamma_i$. This mechanism takes advantage of the randomness introduced by the hash functions. When we vote on arbitrary $L$ beams of the $i$-th RIS, the votes will be distributed over multiple directions, and the direction with the highest number of votes will approximately follow a uniform distribution. However, we can identify the correct direction with a higher probability if the demultiplexing algorithm accurately identifies the signal reflected by the $m_i$-th RIS and then votes on the corresponding set of beam directions $\mathbf{D}^i(\mathbf{q}_{m_i},:)$.
\par
\begin{figure*}[t]\vspace{-2mm}
	\centering
        \subfigcapskip=-3pt
	\subfigure[\textcolor{black}{The first vote.}]{
		\vspace{-1mm}
		\includegraphics[width=0.8\textwidth]{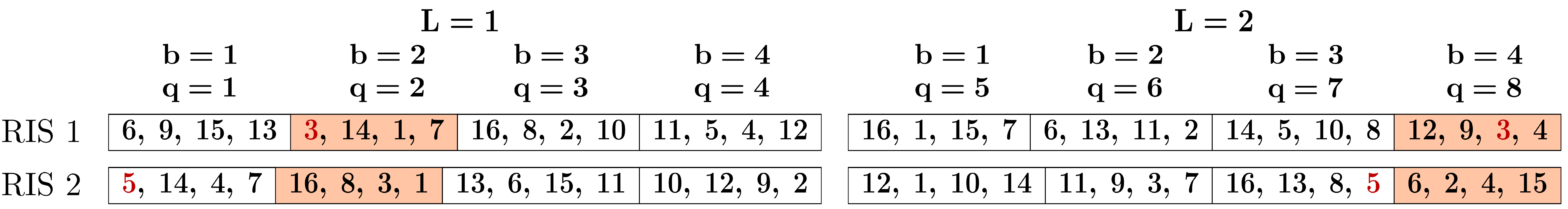}
		\label{fig:vote1}
	}
	\subfigure[\textcolor{black}{The second vote.}]{
		\vspace{-0mm}
		\includegraphics[width=0.8\textwidth]{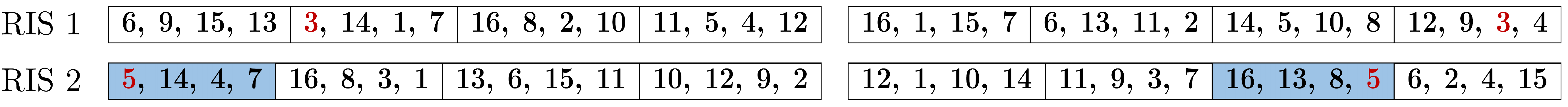}
		\label{fig:vote2}
	}\vspace{-3mm}
	\caption{\textcolor{black}{Proposed multi-RISs voting mechanism with $N=16$, $B=4$, $R=4$.}}\label{fig:vote}\vspace{-4mm}
\end{figure*}

Fig.~\ref{fig:vote} illustrates an example of the voting process in a scenario with two RISs assisting the communication. The time slots with the first $L$ highest received signal power are highlighted in orange, while the time slots with the $L+1$-th to $2L$-th highest power are highlighted in blue. When we vote on the multi-arm beams highlighted in orange in Fig.~\ref{fig:vote1}, we observe that direction $3$ receives the highest number of votes in RIS $1$, while the votes in RIS 2 are more scattered. Consequently, we have $m_1=1$ and $\gamma_1=3$. Subsequently, we vote on the multi-arm beams highlighted in blue of the remaining RIS in Fig.~\ref{fig:vote2}, obtaining $m_2=2$ and $\gamma_2=5$.
\par
The beam identification phase is described in detail in \textbf{Algorithm \ref{alg:Framwork}}. Firstly, the demultiplexing process consists of sorting the received signal power vector $\mathbf{P}^r$ of each user in descending order and obtaining the rearranged time slot sequence numbers $\tilde{\mathbf{q}}$. The number of iterations for the demultiplexing process is determined by the number of RISs, denoted as $I$. However, if the number of RISs is unknown, a threshold value $\epsilon$ can be set, typically equal to the noise power. The process will finish when the remaining received signal power value $P^r(q)$ falls below $\epsilon$, indicating that the signal power is too weak to be reliable. For the $i$-th iteration, the result obtained for RIS $j$ is used to vote on the directions contained in $\mathbf{D}^j(\mathbf{q}_{m_i},:)$, where $\mathbf{q}_{m_i}=\tilde{\mathbf{q}}((i\!-\!1)L+1\!:\!iL)$. Then we obtain the direction $\eta(j)$ with the highest number of votes $\zeta(j)$ among all possible directions and store the voting results as a vector $\bm{\zeta}=\{\zeta(j)\}_{j=1}^I$. The RIS with the highest number of votes can be identified as $m_i=\arg\max\limits_{j\in\{1,\ldots,I\}} \zeta(j)$, and the corresponding direction $\gamma_{m_i}=\eta(m_i)$ represents the user's direction with respect to RIS $m_i$.
\par
\begin{algorithm}[!t]
	\caption{Beam Identification Phase}
	\label{alg:Framwork}
	\begin{algorithmic}[1]
		\Require 
		\Statex the received signal power vectors for all users $\{\mathbf{P}^r_k\}_{k=1}^K$
		\Statex the multi-arm beam direction matrices for all RISs $\{\mathbf{D}^i\}_{i=1}^I$
		\Statex the number of RISs $I$ or the signal power threshold $\epsilon$
		\Statex the number of hashing rounds $L$
		\Ensure 
		\Statex the directions of users corresponding to RISs $\{\bm{\gamma}_k\}_{k=1}^K$, $\bm{\gamma}_k=[\gamma_1,...,\gamma_I]$
		\Statex (For $\forall$ user $k$)
		\State sort the vector $\mathbf{P}^r_k$ in descending order
		\State obtain the rearrangement of the slot sequence $\tilde{\mathbf{q}}$
		\State initialize $i=1$
		\While{$P^r(iL)>\epsilon$}
		\State $\mathbf{q}_{m_i}=\tilde{\mathbf{q}}((i\!-\!1)L+1\!:\!iL)$
			\For{$j$ = 1 to $I$}
			\State $({\eta}(j),{\zeta}(j))\gets$ vote on $\mathbf{D}^j(\mathbf{q}_{m_i},:)$
			\EndFor
			\State $m_i\gets\arg\max\limits_{j\in\{1,\ldots,I\}} {\zeta}(j)$
			\State $\gamma_{m_i}\gets\eta(m_i)$
			\State $i=i\!+\!1$
		\EndWhile
	\end{algorithmic}
\end{algorithm}\setlength{\textfloatsep}{0.3cm}
\par
\textcolor{black}{It is worth noting that the training complexity of the proposed beam training method is $O(B\mathrm{log}N)$, which offers significant advantages. Due to space constraints, the specific analysis is presented in the supplementary material. It enables multiple RISs to scan simultaneously, reducing the overhead of the alternating training method to $1/I$, where $I$ represents the number of RISs involved. Moreover, it uses multi-arm beams, reducing the number of beams required to cover the beam space.}

\vspace{-7.mm}
\section{Simulation Results}\label{sec:simulation_result}
\vspace{0.9mm}
In this section, we present the simulation results of the beam training performance using the proposed beam training method. The number of users is set to $K = 3$. The number of RISs is $I = 3$, which are equipped with $N_v = 32$, $N_h = 32$ reflective elements. The BS consists of an antenna array with $N_A = 64$ antennas. The antenna spacing for both the RIS and BS arrays is set to $d=d_h=d_v=\frac{\lambda}{2}$.
\begin{figure}\vspace{-3mm}
	\begin{center}
		\centerline{\includegraphics[scale=0.33]{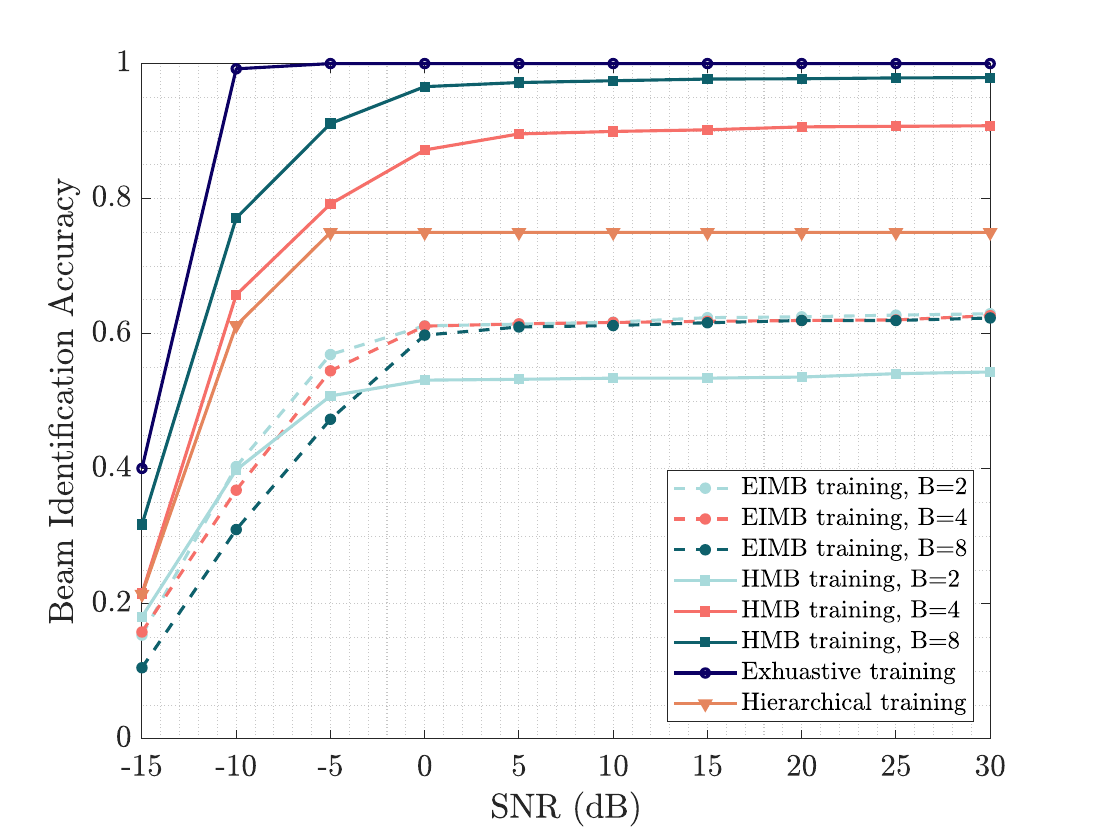}}  \vspace{-2mm}
		\caption{Beam identification accuracy versus the SNR for different training methods when $N=32$. }
		\label{fig:SNR} 
	\end{center}\vspace{-9mm}
\end{figure} 
\begin{figure}
	\begin{center}
		\centerline{\includegraphics[scale=0.33]{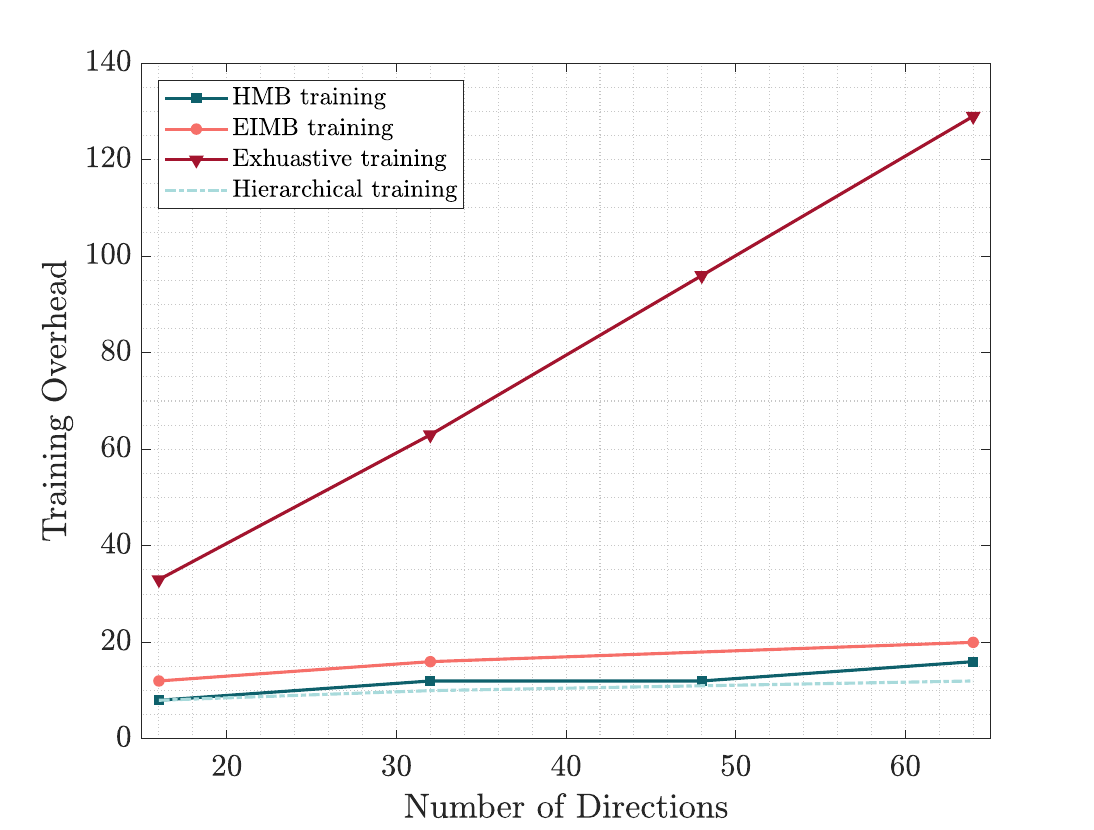}}  \vspace{-2mm}
		\caption{Beam training overhead.}
		\label{fig:time} 
	\end{center}\vspace{-6mm}
\end{figure} 
\par
Fig.~\ref{fig:SNR} plots the effect of the SNR on the beam identification accuracy when the number of directions is fixed at $N=32$. It shows that as the SNR increases, the influence of the noise diminishes, resulting in the increased identification accuracy for all four methods. When the SNR exceeds 0 dB, the accuracy converges. Notably, our proposed method consistently outperforms the other three methods when the number of beams $B\geq 2$, especially at low SNRs. For example, with $B = 8$, our proposed method achieves an accuracy of approximately 97.5\%, achieving a significant improvement of at least 20\% compared to existing
methods. 
\par
Fig.~\ref{fig:time} plots the relationship between the number of directions and the training overhead. We fixed the SNR to 5 dB and the identification accuracy to 60\%. While the exhaustive beam training, hierarchical beam training and the EIMB training methods train alternately, our proposed HMB training method trains simultaneously, with the complexity not increasing with the increase in the number of RISs or users. It can be seen that the complexity of the HMB training method is on the logarithmic level, which greatly reduces the training overhead of the traditional methods. It is worth noting that although the hierarchical training method has a lower complexity at an accuracy of 60\%, it was shown in the last figure that this method limits the accuracy to 75\%.

\vspace{-1.8mm}
\section{Conclusion}\label{sec:conclusion}
\vspace{1.2mm}
In this paper, a HMB training method is proposed for multi-RIS-assisted multi-user communication systems. The proposed method utilizes independent hash functions to generate multi-arm beams. It effectively demultiplexes the signals reflected from different RISs by employing the soft decision on the received signal power. Further, we design a multi-round voting mechanism to obtain the aligned direction. Simulation results demonstrate the robustness and effectiveness of our proposed method in beam identification accuracy. Compared to existing methods, our approach achieves a significant improvement in accuracy of at least 20\%. Furthermore, our method ensures that the training overhead remains manageable even with an increasing number of RISs and users, as it remains at the logarithmic level.

\vspace{0mm}
\bibliographystyle{IEEEbib}
\bibliography{strings}
\end{document}


\title{\Large\textbf{Supplementary Material of Low-Complexity Beam Training for Multi-RIS-Assisted Multi-User Communications}\vspace{-20mm}}
\maketitle

\thispagestyle{empty}
\pagestyle{empty}
\section{hash function generation method}
Define the universe of keys and the hash values: We denote the universe of keys as the collection of all codeword indexes $\mathcal{U}=\{0,1,...,N_C-1\}$. We define the interpreted hash values as $\mathcal{T}=\{0,1,...,B-1\}$, where a hash function $f:\mathcal{U}\to\mathcal{T}$ maps each key (codeword index) to the interpreted hash value within the desired range. Randomly select hash function $f$ from a family $\mathcal{H}$. The family $\mathcal{H}=\{h_1,h_2,...,h_{|\mathcal{H}|}\}$ consists of distinct hash functions, and $|\mathcal{H}|$ represents the total number of distinct functions in the family. The construction of the family $\mathcal{H}$ is based on the following theorem
\begin{theorem}
For the Galois field $GF(N_C)$, we construct a $k$-wise independent map from $\mathcal{U}$ to $\mathcal{T}$ as follows:
\par
Pick $k$ random numbers $a_0,a_1,...,a_{k-1}\in GF(N_C)$. For any $x\in\mathcal{U}$, 
\begin{equation}
    h(x) \triangleq a_0 + a_1x + a_2x^2 + ... + a_{k-1}x^{k-1}\ \rm{mod}\ B,
\end{equation}
where the calculations are done over the field $GF(N_C)$, and $|\mathcal{H}|=N_C^{k}-N_C$, because $a_1,a_2,... . a_{k-1}$ cannot equal to zero at the same time.
\end{theorem}
\textit{Proof:}
Define 
\begin{equation}
    \begin{bmatrix}
        1 & x_1 &x_1^2&...&x_1^{k-1} \\
        1 & x_2 &x_2^2&...&x_2^{k-1} \\
        ...\\
        1 & x_k &x_k^2&...&x_k^{k-1}
    \end{bmatrix}
    \begin{bmatrix}
        a_0 \\
        a_1 \\
        a_2 \\
        ...\\
        a_{k-1}
    \end{bmatrix}
    =
    \begin{bmatrix}
        r_1 \\
        r_2 \\
        ...\\
        r_k
    \end{bmatrix}.
\end{equation}
\par
Thus,
\begin{equation}
    \begin{bmatrix}
        a_0 \\
        a_1 \\
        a_2 \\
        ...\\
        a_{k-1}
    \end{bmatrix}
    =
    \begin{bmatrix}
        1 & x_1 &x_1^2&...&x_1^{k-1} \\
        1 & x_2 &x_2^2&...&x_2^{k-1} \\
        ...\\
        1 & x_k &x_k^2&...&x_k^{k-1}
    \end{bmatrix}^{-1}
    \begin{bmatrix}
        r_1 \\
        r_2 \\
        ...\\
        r_k
    \end{bmatrix},
\end{equation}
where the matrix is invertible because for $\forall i\neq j \in \{1,...,k\}$, $x_i\neq x_j$, and matrix inverses are done over the finite field $GF(N_C)$. This implies that when fixing $x_1, x_2,...,x_k$, there is only a unique set of coefficients $a_0,a_1,...,a_{k-1}$ corresponding to $r_1,r_2,...,r_k$, as     \begin{equation}\label{eq:pr1}
        \begin{split}
        &Pr(h(x_1)=\alpha_1\ \wedge\ h(x_2)=\alpha_2\ \wedge\ ...\ \wedge\ h(x_k)=\alpha_k)\\
        =Pr(r_1\ \rm{mod}& \ B=\alpha_1\ \wedge\ r_2\ \rm{mod}\ B=\alpha_2\ \wedge\ ...\ \wedge\ r_k\ \rm{mod}\ B=\alpha_k), \ \ \ \alpha_{i}\neq\alpha_{j}\in\mathcal{T}.
        \end{split}
    \end{equation} 
    The mentioned polynomial serves as a hash function, mapping the elements from the universe $\mathcal{U}$ to itself. To form a family of hash functions that map $\mathcal{U}\to\mathcal{T}$, we discard $\lceil \log_2 N_C\rceil-\lceil \log_2B\rceil$ bits from the values $r_i$. It results in a collection of hash functions that effectively map elements from the universe $\mathcal{U}$ to the interpreted hash values within the designated range $\mathcal{T}=\{0, 1, ..., B-1\}$.
    \par

\section{Complexity Analysis}
\par
The number of hash rounds is only related to the desired identification accuracy. As demonstrated by \textbf{Theorem \ref{complexity-2}}, which proves that as long as the number of hash rounds $L=O(\mathrm{log} M_s)$, the identification error is about $1/M_s$. That is, we sacrifice some accuracy compared to the exhaustive approach but greatly reduce the training overhead to $O(B\mathrm{log}N)$, where $B$ is the number of multi-arm beams, and $N$ is the total number of directions.

\begin{theorem}\label{complexity-2}
Given the number of hash rounds is $L=O(\mathrm{log}M_s)$, it is guaranteed that the probability of identification error is less than $\frac{1}{M_s}$.
\end{theorem}
\par\vspace{1mm}
\textit{Proof:}
\begin{lemma}
(Hoeffding's Lemma) For random variables $X$, $P(X\in[a,b])=1$, $E[X]=0$, there are
\begin{equation}
	\begin{split}
		\forall s,\quad E[e^{sX}]\leq e^{(s^2 (b-a)^2/8)}.
	\end{split}
\end{equation}
\end{lemma}
\par
Firstly, for the $l\in\{1,...,L\}$-th round of hash, the power of the signal received by the $b$-th multi-arm beam is denoted as $P(l,b)$, where $q=(l-1)B+b$ represents the time slot number. Since the hash functions for the $L$ rounds of hash are randomly chosen from the hash function family, the values in $\mathcal{L}_b=\{P(1,b),...,P(l,b),...,P(L,b)\},\forall b$, are random variables. The number of elements in set $\mathcal{L}_b$ is denoted as $card(\mathcal{L}_b)=L_b$. Additionally, these variables are independent due to the $k$-wise independence property.  
\par
Furthermore, since the received signal power values are positive and finite, we have $P(l,b)\in[P_{min},P_{max}], l\in\{1,...,L\}$, where $P_{min}$ and $P_{max}$ represent the upper and lower bounds of the power, respectively. This leads to 
\begin{equation}
    \begin{split}
        \forall s,t>0,\\
        &Pr(P(l,b)-E[P(l,b)]\geq t)=Pr(e^{s(P(l,b)-E[P(l,b)])} \geq e^{st})\\
        &\overset{(b)}{\leq} e^{-st}E[e^{s(P(l,b)-E[P(l,b)])}]=e^{-st}\prod\limits_{l=1}^L E[e^{s\frac{P(l,b)-E[P(l,b)]}{L}}]\\
        &\overset{(c)}{\leq} e^{-st}\prod\limits_{l=1}^L E[e^{s^2(\frac{P_{max}-P_{min}}{L})^2/8}]=e^{-st+\frac{s^2}{8}\sum\limits_{l=1}^{L}(\frac{P_{max}-P_{min}}{L})^2}\\
        &=e^{-st+\frac{s^2(P_{max}-P_{min})^2}{8L}}.
    \end{split}
\end{equation}
The inequality $(b)$ is based on Markov's inequality because the random variable is non-negative and has a finite mean, which satisfies the preconditions of Markov's inequality, $(c)$ is based on Hoeffding's Lemma. Since the index $b$ can take the value from $1$ to $B$, and each multi-arm beam is equivalent due to the randomness of the hash function, we can use the same threshold $E[P(l,b)]\triangleq T_0 $. Then we have 
\begin{equation}
    \frac{P(l,b)-E[P(l,b)]}{L}\in[\frac{P_{min}-T_0}{L},\frac{P_{max}-T_0}{L}],
\end{equation}
\begin{equation}
    E[\frac{P(l,b)-E[P(l,b)]}{L}]=0.
\end{equation}
\par
To obtain the best probability upper bound, we minimize the right-hand side of the equation concerning $s$, define
\begin{equation}
    g(s)=-st+\frac{s^2 (P_{max}-P_{min} )^2}{8L}=-st+\frac{s^2(\Delta P)^2}{8L}.
\end{equation}
\par
This equation is a quadratic function that takes the minimum value when $s=\frac{4tL}{(\Delta P)^2}$, thus\vspace{-1mm}
\begin{equation}
    g(s)_{min}=-\frac{2t^2L}{(\Delta P)^2},
\end{equation}
where $\Delta P\triangleq  P_{max}-P_{min}$.
\par\vspace{2mm}
Consequently, we have\vspace{-1mm}
\begin{equation}
	\begin{split}\label{eq:eq11}
		\forall t>0,\forall b,\quad Pr(P(l,b)-E[P(l,b)]\geq t) \leq e^{-\frac{2t^2L}{(\Delta P)^2}}.
	\end{split}
\end{equation}
\par\vspace{1mm}
Because $P(l,b)-E[P(l,b)]=-(E[P(l,b)]-P(l,b))$, in the same way we can obtain
\vspace{-2mm}
\begin{equation}
	\begin{split}\label{eq:eq22}
		\forall t>0,\forall b,\quad Pr(E[P(l,b)]-P(l,b)\geq t) \leq e^{-\frac{2t^2L}{(\Delta P)^2}}.
	\end{split}
\end{equation}
\par\vspace{1mm}
Divide the random variable $P(1,b),...,P(l,b),...,P(L,b)$ into those that contain user signals and those that do not (i.e., only the noise), denote as $\mathcal{L}_b^{s}=\{P(l,b)^{(s)}\}$ and $\mathcal{L}_b^{ns}\{P(l,b)^{(ns)}\}$, respectively. We have\vspace{1mm} 
\begin{equation}
    \mathcal{L}_b^{s}\in[P_{min}^{s},P_{max}^{s}],\quad E[\mathcal{L}_b^{s}]\triangleq T_b^{s},
\end{equation}
\begin{equation}
    \mathcal{L}_b^{ns}\in[P_{min}^{ns},P_{max}^{ns}],\quad E[\mathcal{L}_b^{ns}]\triangleq T_b^{ns},\vspace{1mm}
\end{equation}
where $P_{min}^{s},P_{max}^{s}$ are the lower and upper bounds of the received power of the useful signal, $P_{min}^{ns},P_{max}^{ns}$ are the lower and upper bounds of the received power of the noise. \vspace{1mm}
\begin{equation}
\begin{split}
    card(\mathcal{L}_b^{s})&=L_b^{s},\quad card(\mathcal{L}_b^{ns})=L_b^{ns},\\
    &L_b^{s}+L_b^{ns}=L_b.
\end{split}
\end{equation}
\par
Since 
\begin{equation}
    T_b^{s} L_b^{s}+T_b^{ns} L_b^{ns}=T_0L_b,
\end{equation}
we have $T_b^{ns}<T_0<T_b^{s}$. 
\par
According to Eq.~(\ref{eq:eq11}), if we take $t=T_0-T_b^{ns}>0$ and 
\begin{equation}
    L^{ns}=\frac{\mathrm{ln}M_s (P_{max}^{ns}-P_{min}^{ns})^2}{2(T_0-T_b^{ns})^2},
\end{equation}
we have \vspace{-2mm}
\begin{equation}
	\begin{split}
		Pr(P(l,b)^{(ns)}\geq T_0 )\leq e^{-\frac{2(T_0-T_b^{ns})^2 L^{ns}}{(P_{max}^{ns}-P_{min}^{ns})^2 }}=\frac{1}{M_s}.
	\end{split}
\end{equation}
\par
Similarly, according to Eq.~(\ref{eq:eq22}), if we take $t=T_b^{s}-T_0>0$ and
\begin{equation}
    L^{s}=\frac{\mathrm{ln}M_s (P_{max}^{s}-P_{min}^{s})^2}{2(T_0-T_b^{s})^2},
\end{equation}
we have \vspace{-2mm}
\begin{equation}
	\begin{split}
		Pr(P(l,b)^{(s)}\leq T_0 )\leq e^{-\frac{2(T_0-T_b^{s})^2 L^{s}}{(P_{max}^{s}-P_{min}^{s})^2 }}=\frac{1}{M_s},
	\end{split}
\end{equation}
\begin{equation}\label{eq:LL}
    L^{ns}=O(\mathrm{log}M_s),\quad L^{s}=O(\mathrm{log}M_s).
\end{equation}
\par
Eventually, the probability of identification error is reduced to $\frac{1}{M_s}$ when Eq.~(\ref{eq:LL}) holds, we obtain
\begin{equation}
    L=L^s+L^{ns}=O(\mathrm{log}M_s).
\end{equation}